\documentclass[a4paper,10pt]{article}
\usepackage{here}
\usepackage{amsmath}
\usepackage{mathtools}
\usepackage{graphicx}
\usepackage{color}
\usepackage{tikz}
\usepackage{authblk}
\title{ Wetting of Ga droplets in SiO$_2$/Si cavities: Application to self-assisted GaAs nanowire growth}
\author[1,2]{Louis Bailly-Salins}
\author[1,3]{Marco Vettori}
\author[1]{Thomas Dursap}
\author[1]{Philippe Regreny}
\author[4]{Gilles Patriarche}
\author[1]{Michel Gendry}
\author[1]{José Penuelas}
\author[1]{Alexandre Danescu}
\affil[1]{University of Lyon, Institute of Nanotechnologies de Lyon - INL, UMR CNRS 5270, Ecole Centrale de Lyon, 69130 Ecully, France}
\affil[2]{present address : Laboratoire de Physique Corpusculaire, ISMRA, IN2P3-CNRS et Université de Caen, 14050 Caen Cédex, France}
\affil[3]{present address : Department of Applied Physics, Eindhoven University of Technology, Groene Loper 19, 5612AP Eindhoven, The Netherlands}
\affil[4]{Centre de Nanosciences et de Nanotechnologies - C2N, CNRS, Univ. Paris-Sud, Université Paris-Saclay, 91120, Palaiseau, France}
\date{}

\begin{document}
\maketitle

\begin{abstract}
In this paper we compute and compare the surface energy of various Ga liquid droplets wetting a cylindrical cavity in various configurations. While for some of these configurations the surface energy can be computed explicitely for others numerical computation is needed. Motivated by the results obtained for the cylindrical cavities we explore the case of the more realistic situation, conical cavities. Our results provide a relation between the geometry of the conical cavity and the equilibirum wetting angles of the droplet on the bottom and on the sidewall of the cavity which insure the dewetting of the lateral surface. This is an important result toward the control of the verticality during the nanowire growth by the vapor liquid solid method.
\end{abstract}

\section{Introduction}
Nanowires (NWs) are promising building blocks for many functional devices thanks to their exceptional properties \cite{MAC20}. The most popular NW growth method is based on the Vapor-Liquid-Solid (VLS) mechanism \cite{WAG64}. The VLS mechanism consists in the formation of a solid phase due to the saturation of liquid droplets which are exposed to a vapor phase. The droplets are usually randomly deposited on a solid surface resulting in a random organization of the NWs \cite{FOU19}. However, in order to benefit from collective effects like in photonic crystals \cite{RA17} it is necessary to fabricate periodic arrays of NWs. Moreover achieving patterned growth of NWs is a good way to optimize their size distribution and their physical properties \cite{VUK17}. As a consequence, a key issue for many applications is to fabricate organized patterns of particle catalyst nanoparticles on the surface that will act as nucleation seeds for the VLS NW growth. Although the fabrication of such periodic arrays of metallic nanoparticles has been investigated from years \cite{KIN08,DEM21,VUK19}, it is still a challenge to achieve low cost and large area patterns. These patterns are usually fabricated by lithography techniques which allow to obtain periodic arrays of holes on a surface. Then under appropriate experimental conditions the liquid droplets can be deposited in the holes and be exposed to the vapor phase. However, experimental results reported in the literature highlight the complexity of the NW nucleation when the liquid droplet is confined in patterned holes \cite{VUK19}. In particular, one can assume that the shape of the liquid droplet, its surface energy \cite{YUA18} and the wetting on the hole surfaces have a strong impact of the first steps of the crystal nucleation. Although recent experimental advances in characterization methods such as in situ TEM allow to provide a better picture of the VLS mechanism \cite{HAR18, MAL21}, a clear description of the mechanism of NW nucleation in a cavity is still missing. Moreover, only few studies reported theoretical explanation of the wetting of liquid in holes with geometry corresponding to those of patterned substrates. 
In this article, we studied the wetting of a liquid droplet in cylindrical and conical cavities. Various wetting configurations as well as hole geometries were considered and their corresponding surface energy was calculated in order to give a clear picture of the phenomena involved during the first steps of the VLS growth on patterned substrates. Applied to the particular case of a Ga droplet in a SiO$_2$/Si conical cavity, the results obtained are in excellent agreement with our experiments \cite{VET19} and pave the way to a better control of the NW verticality  on patterned substrates \cite{MAT16}.

\section{Surface energy and droplet configurations}
The presence of the droplet in the cylindrical cavity replaces some of solid/vapor interfaces by solid/liquid interfaces and creates a new liquid/vapor interface. Thus, it seems natural to associate to a given morphology of the droplet the corresponding variation of the surface energy, further denoted $\Delta W,$ and defined as the surface energy of the system in the presence of the droplet minus the reference surface energy of the cavity without the droplet. This gives
\begin{equation}
\Delta W = A_{VL} \gamma_{VL} + A_{LS}(\gamma_{LS}-\gamma_{SV}),
\label{generic}
\end{equation}
where $A_{LV}$ and $A_{LS}$ are the areas of the vapor/liquid and liquid/solid interfaces, respectively and $\gamma_{VL}, \gamma_{LS}$ and $\gamma_{SV}$ are the surface energies of the vapor/liquid, the liquid/solid and the solid/vapor interfaces, respectively. In (\ref{generic}) the first term accounts for the new created vapor/liquid interface and the second term accounts for the area of the liquid/solid interface that replaces various solid/vapor (bottom and/or wall) interface. 

For generality, we shall consider here the generic situation in which the wetting angles of the droplet on the bottom ($\theta_b$) and the walls ($\theta_w$) of the cavity respectively, are different. This is for instance the case when one discuss liquid droplets in cavities formed in a thin $\textrm{SiO}_2$ layer on cristalline $\textrm{Si}.$ Using the Young relation for each of the solid/vapor interface we can rewrite (\ref{generic}) as
\begin{equation}
\Delta W = \gamma_{VL}\left(A_{VL} -  A_{LS}^{(b)}\cos \theta_b -  A_{LS}^{(w)}\cos\theta_w\right),
\label{general}
\end{equation}
where $A_{LS}^{(b)}$ denotes the area of the liquid/solid interface located at the bottom of the cavity and $A_{LS}^{(w)}$ the area of the liquid/solid interface between the droplet and the wall. For later references $A_{LS}=A_{LS}^{(b)}+A_{LS}^{(w)}$ denotes the total area of the solid/drop interface. Notice that our definition in (\ref{general}) is independent on the droplet morphology and satisfies $\Delta W = 0$ when the droplet is absent.

\begin{figure}[ht!]
    \centering
    \includegraphics[width=\textwidth]{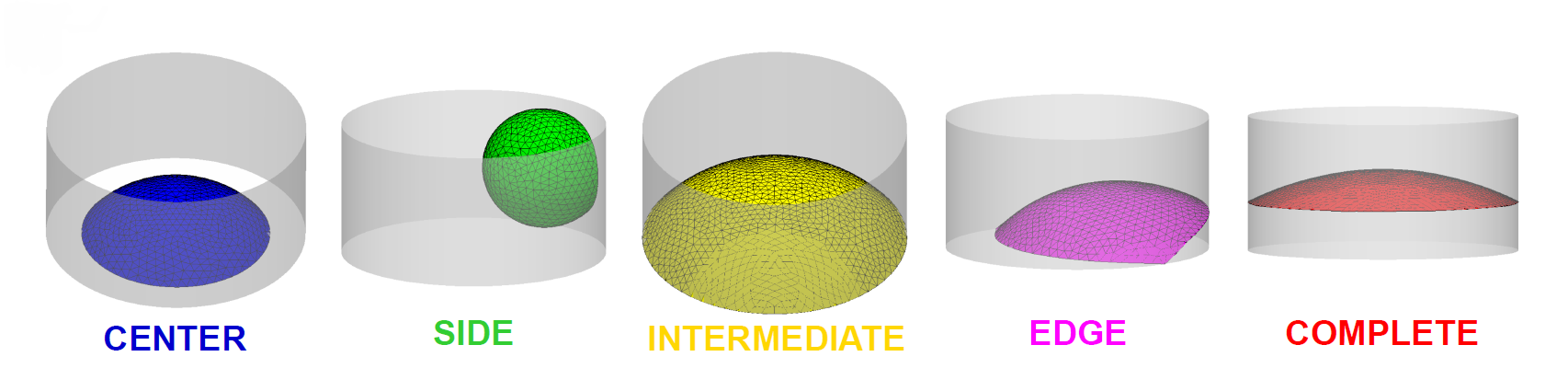}
    \caption{The liquid/vapor interface for morphological different configurations of a droplet in a cylindrical cavity. From left to right : CENTER, SIDE, INTERMEDIATE, EDGE and COMPLETE. The color code adopted here will be conserved for all subsequent graphics.}
    \label{fig:config}
\end{figure}

Depending on the position of the droplet in the cavity, using notations in \cite{VUK19}, we define several different configurations (Figure 1 illustrates these in the case of the cylindrical cavity):
\begin{itemize}
\item In a CENTER configuration the droplet wets only the bottom of the cavity at wetting angle $\theta_b.$
\item In a SIDE configuration the droplet wets only the wall of the cavity at wetting angle $\theta_w.$
\item In an INTERMEDIATE configuration all of the bottom of the cavity is wet but not the wall, so that the droplet is pinned at the line between the bottom and the wall. In this situation (see Gibbs \cite{gibbs}) the droplet has a volume dependent contact angle $\theta_i$ such that
$$\min(\theta_b,\theta_w-\frac{\pi}{2})\leq \theta_i \leq \max(\theta_b,\theta_w-\frac{\pi}{2}).$$
\item In an EDGE configuration the droplet wets partially both the bottom and the wall of the cavity.
\item In a COMPLETE configuration all of the bottom of the cavity is wet and the triple line lies strictly above the cavity basis.
\end{itemize}
Obviously, the INTERMEDIATE configuration exists only if $\max(\theta_b,\theta_w-\frac{\pi}{2})>\min(\theta_b,\theta_w-\frac{\pi}{2})$ and can be seen as a transition between the CENTER and COMPLETE configurations. In particular, if $\theta_b=\theta_w,$ the INTERMEDIATE configuration exists for an angle interval of $\pi/2.$ Clearly, all configurations defined above exist only for (relative) volumes in specific intervals described below.

\section{Droplet in a cylindrical cavity}

Given the surface tensions or equivalent, the contact angles $\theta_b$ and $\theta_w$, by using elementary geometry, one can compute explicitly $\Delta W$ for the three of the above configurations: CENTER, INTERMEDIATE, and COMPLETE. Indeed, for the CENTER and INTERMEDIATE configurations the droplet shape of minimum surface energy is that of a spherical cap, while for the COMPLETE configuration a spherical cap placed on a cylinder. Thus its area is known, so its surface energy.

We shall discuss first the case $\theta_b<\pi/2$ and $\theta_w>\pi/2$ (which is that of a cylindrical cavity in an $\textrm{SiO}_2$ layer on a $\textrm{Si}$ substrate). In this case we have :

\begin{itemize}
\item The CENTER configuration: this is a spherical cap and exists only if the volume of the droplet $V$ is such that  $0<V<V(R,\theta_b)$ where the function $$V(r,\theta) =\frac{\pi r^3(1-\cos\theta)^2(2+\cos\theta)}{3\sin^3\theta},$$ gives the volume of a spherical cap with basis radius $r$ and wetting angle $\theta.$ In this configuration we have
    \begin{eqnarray}
    \frac{\Delta W}{\gamma_{VL}} & = & A_{VL}-cos\theta_b A_{LS}=\nonumber \\
    &=& A_{VL}(1-cos\theta_b\frac{1+\cos\theta_b}{2})=\nonumber \\
    & = & 2\left(\frac{3V}{2+\cos\theta_b}\right)^{2/3}\left(\frac{\pi}{1-\cos\theta_b}\right)^{1/3}\left(2-\cos\theta_b-\cos^2\theta_b\right)\nonumber \\
    & = & V^{2/3}\left[36 \pi (2+\cos\theta_b)\sin^4(\theta_b/2)\right]^{1/3}.
    \label{center}
    \end{eqnarray}
    We notice that $\lim_{V\rightarrow 0} \Delta W = 0$ as expected.
\item The INTERMEDIATE configuration: let $\theta_m = \min(\theta_b,\theta_w-\pi/2)),$ $\theta_M = \max(\theta_b,\theta_w-\pi/2)),$ $V_m = V(R,\theta_m)$ and $V_M=V(R,\theta_M).$ This configuration is also a spherical cap and exists only if the volume of the droplet lies in the interval $[V_m,V_M].$ In this case we have
    \begin{equation}
    \frac{\Delta W}{\gamma_{VL}} = 2\left(\frac{3V}{2+\cos\theta}\right)^{2/3}\left(\frac{\pi}{1-\cos\theta}\right)^{1/3}-\pi R^2\cos\theta_b,
    \label{inter}
    \end{equation}
    where $\theta$ is the unique positive solution in the interval $[\theta_m,\theta_M]$ of the equation $V(R,\theta)=V.$
\item The COMPLETE configuration is a spherical cap on a cylinder and exists if the droplet volume lies in the interval $[V_M,V_M+\pi hR^2].$
    In this case
    \begin{equation}
    \frac{\Delta W}{\gamma_{VL}} = 2\left(\frac{3V}{2+\cos\theta_M}\right)^{2/3}\left(\frac{\pi}{1-\cos\theta_M}\right)^{1/3}-\pi R^2\cos\theta_b-2\pi Rh \cos\theta_w.
    \label{complete}
    \end{equation}
\end{itemize}

A straightforward computation shows that $\Delta W$ defined in (\ref{center})-(\ref{complete}) is a continuous function. Let the aspect ratio of the cylindrical cavity be defined as $R/h.$ Figure 2 shows the (normalized) surface energy of CENTER, INTERMEDIATE and COMPLETE configurations as functions of the droplet volume for a cavity with $R=h=1$ (aspect ratio 1) and three qualitatively different values for the pair $(\theta_b,\theta_w),$ i.e. $(45^\circ,105^\circ),$ $(45^\circ,135^\circ)$ and $(45^\circ,155^\circ).$ In the first case the upper limit value of the CENTER configuration is higher than the lower value of the COMPLETE configuration, since we have $\theta_w-\pi/2<\theta_b.$ In the second case the INTERMEDIATE configuration does not exist as we have exactly $\theta_w-\pi/2 = \theta_b.$ In the third case the INTERMEDIATE configuration connect the CENTER configuration to the COMPLETE configuration as $\theta_w-\pi/2>\theta_b.$ The case of a $\textrm{Ga}$ droplet into a $\textrm{SiO}_2/\textrm{Si}$ cylindrical cavity falls in the first class as $\theta_b=50^\circ$ and $\theta_w=116^\circ.$

\vspace*{1.cm}
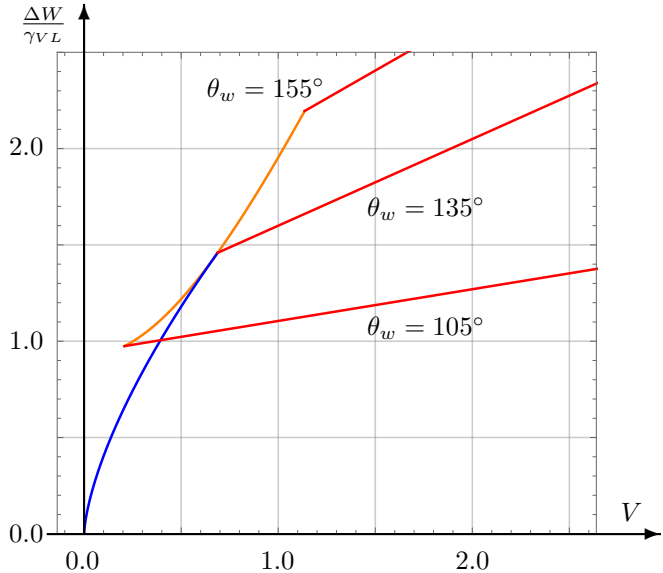
\begin{figure}[ht!]\centering
\includegraphics[width=0.5\textwidth]{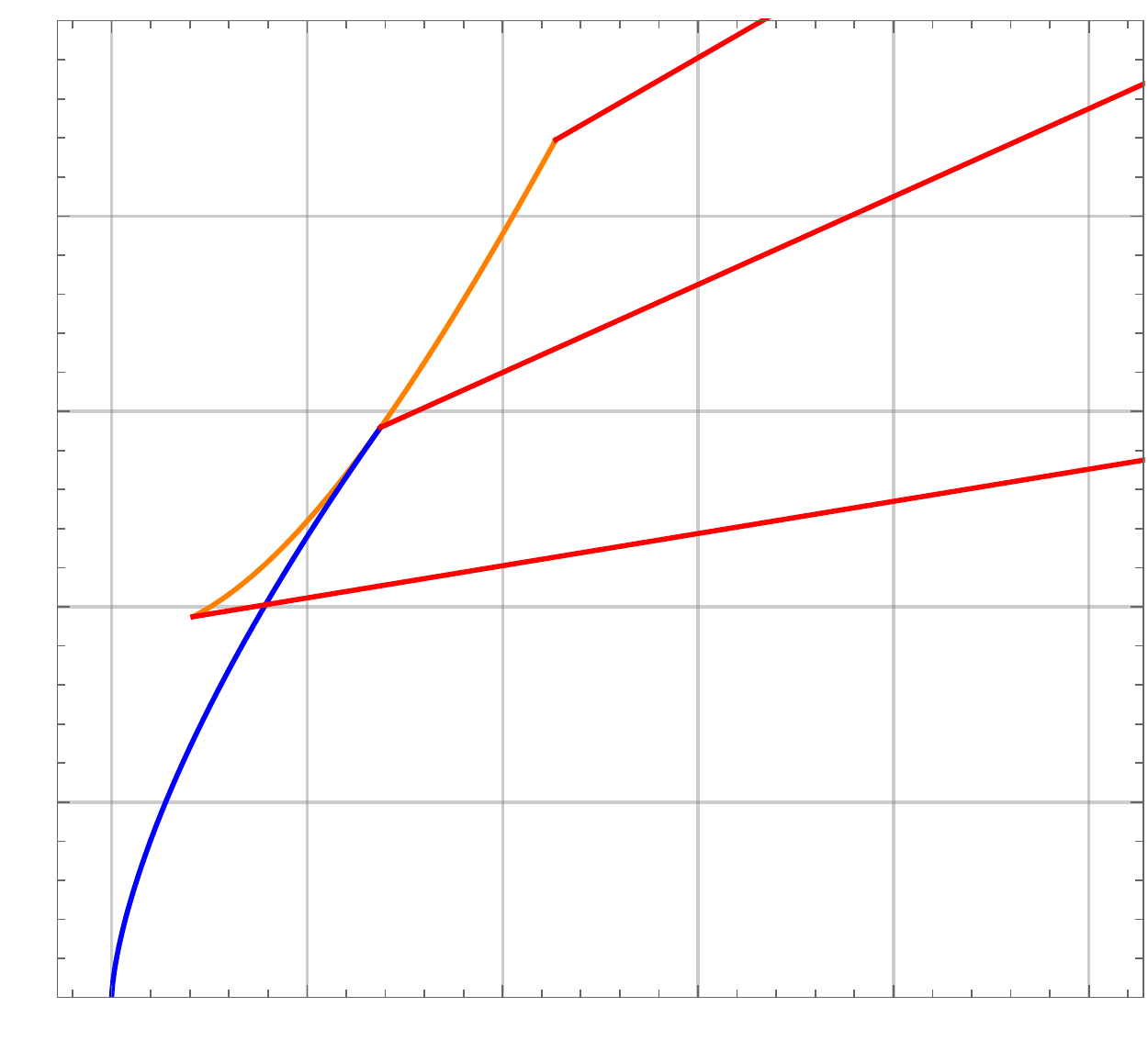}
\begin{picture}(0,0)(0,0)
\put(5,15){$V$}
\put(-220,200){$\frac{\Delta W}{\gamma_{VL}}$}
\put(-90,85){$\theta_w=105^\circ$}
\put(-90,130){$\theta_w=135^\circ$}
\put(-150,175){$\theta_w=155^\circ$}
\thicklines
\put(-210,10){\vector(1,0){230}}
\put(-196,10){\vector(0,1){200}}
\put(-203,-3){$0.0$}
\put(-130,-3){$1.0$}
\put(-57,-3){$2.0$}
\put(-224,153){$2.0$}
\put(-224,80){$1.0$}
\put(-224,7){$0.0$}
\end{picture}
\label{center_inter_etc}
\caption{Illustration of the normalized surface energy for the CENTER (blue), INTERMEDIATE (orange) and COMPLETE (red) configurations for $(\theta_b,\theta_w)=(45^\circ,105^\circ),$ $(45^\circ,135^\circ)$ and $(45^\circ,155^\circ).$ In the second case, since $\theta_m=\theta_M,$ the INTERMEDIATE configuration does not exists. Results obtained for $R=h=1.$}
\end{figure}

For the EDGE and SIDE configurations there are no explicit expressions for the areas that realize the minimum of the surface energy. Consequently, the corresponding surface energies were computed using the Surface Evolver software \cite{site, SE_article}, which minimize the total surface energy of a system subject to geometric and/or energetic constraints (see also \cite{manuel}).

\subsection{Results for a Ga droplet in a SiO${}_2$/Si cylindrical cavity}

Ga droplets wet Si(111) surfaces at wetting angle\footnote{Obviously, these numerical values should be carrefully considered since their precise measurement is difficult. Hereafter, for practical purposes in our numerical investigation, we used these values but, obviously, all our results can be adapted to different numerical data.} $\theta_b=50^\circ$ and $\textrm{SiO}_2$ surfaces at $\theta_w = 116^\circ$. Comparing the normalized surface energies $E_{\rm drop}/(A_{\rm cav}\gamma_{LV})$ obtained for each configuration above allows to determine, as functions of the volume of the droplet and the cavity aspect ratio, the minimum surface energy, and thus the optimal morphology, of the droplet. Figure 3 shows the normalized surface energies as a function of the (normalized) droplet volume in two generic situations: for $R/h=1$ and for $R/h=3.5.$ 

\begin{figure}[ht!]
    \begin{centering}
    \includegraphics[width=\textwidth]{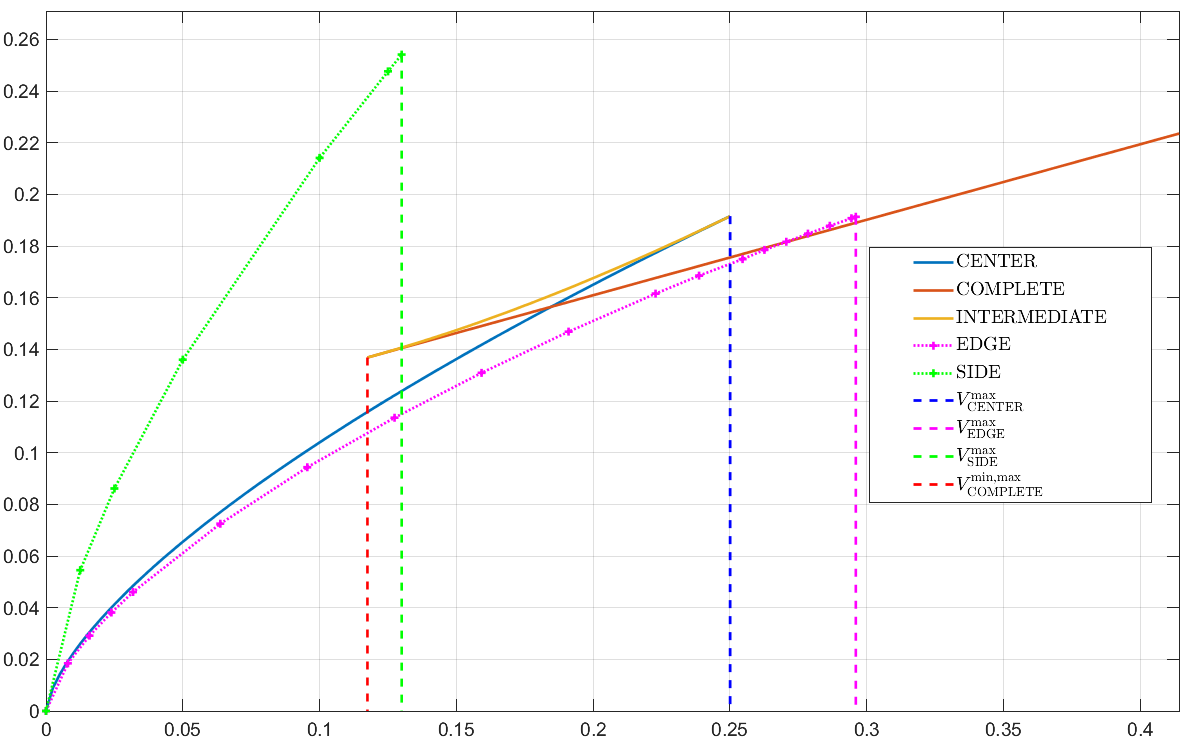}
   \includegraphics[width=\textwidth]{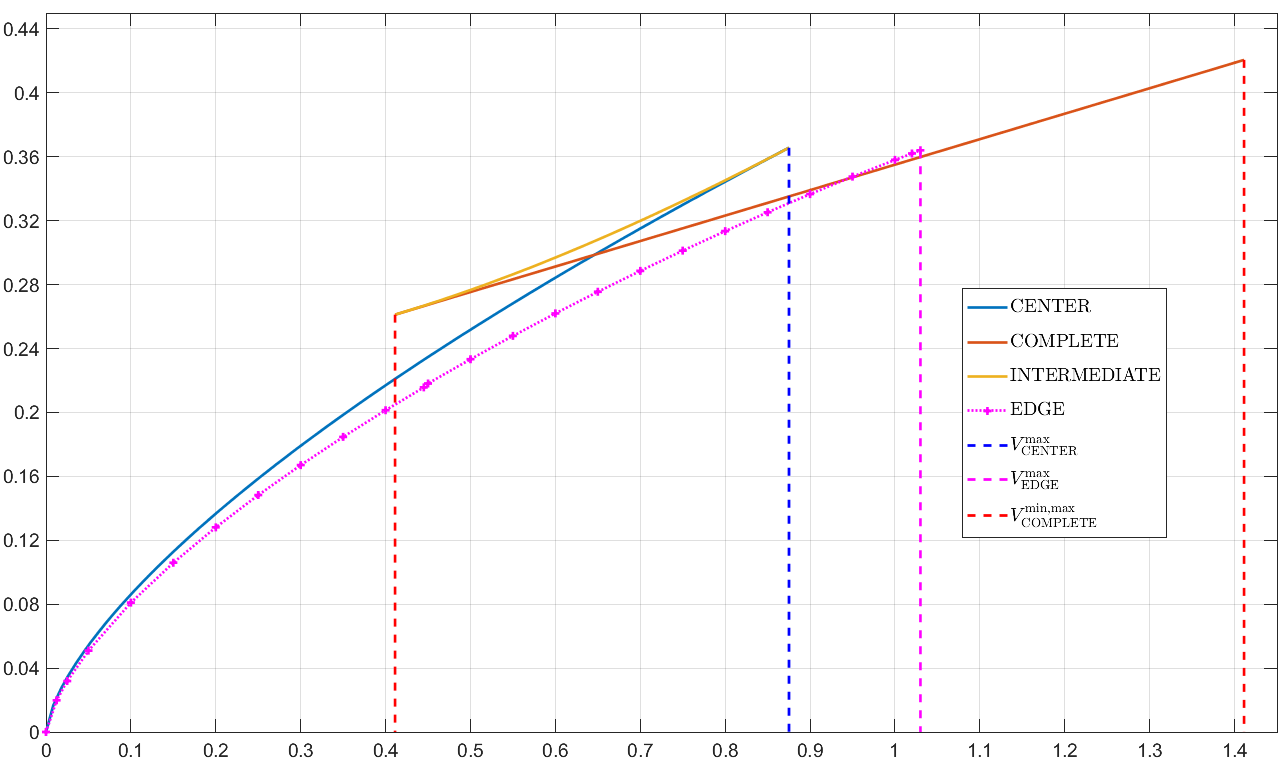}
    \begin{picture}(0,0)
     \put(200,500){$\color{black} {\bf a})$}
     \put(200,220){$\color{black} {\bf b})$}
     \put(30,160){\rotatebox{90}{$\Delta W/(A_{\rm cav}\gamma_LV{})$}}
     \put(30,420){\rotatebox{90}{$\Delta W/(A_{\rm cav}\gamma_LV{})$}}
     \put(360,35){\rotatebox{0}{$V_{\rm drop}/V_{\rm cav}$}}
     \put(360,290){\rotatebox{0}{$V_{\rm drop}/V_{\rm cav}$}}
    \end{picture}
    \end{centering}
    \caption{\textbf{a)} Energies of the different configurations for an aspect ratio $R/h = 1$ as a function of the normalized droplet volume. In full lines: the energies calculated analytically. In dashed lines: the energies for the SIDE and EDGE configurations computed using Surface Evolver. \textbf{b)} Same as (a) but for $R/h = 3.5$. The SIDE configuration energy is not represented here due to its higher energy.}
    \label{fig:wsurh2-7}
\end{figure}
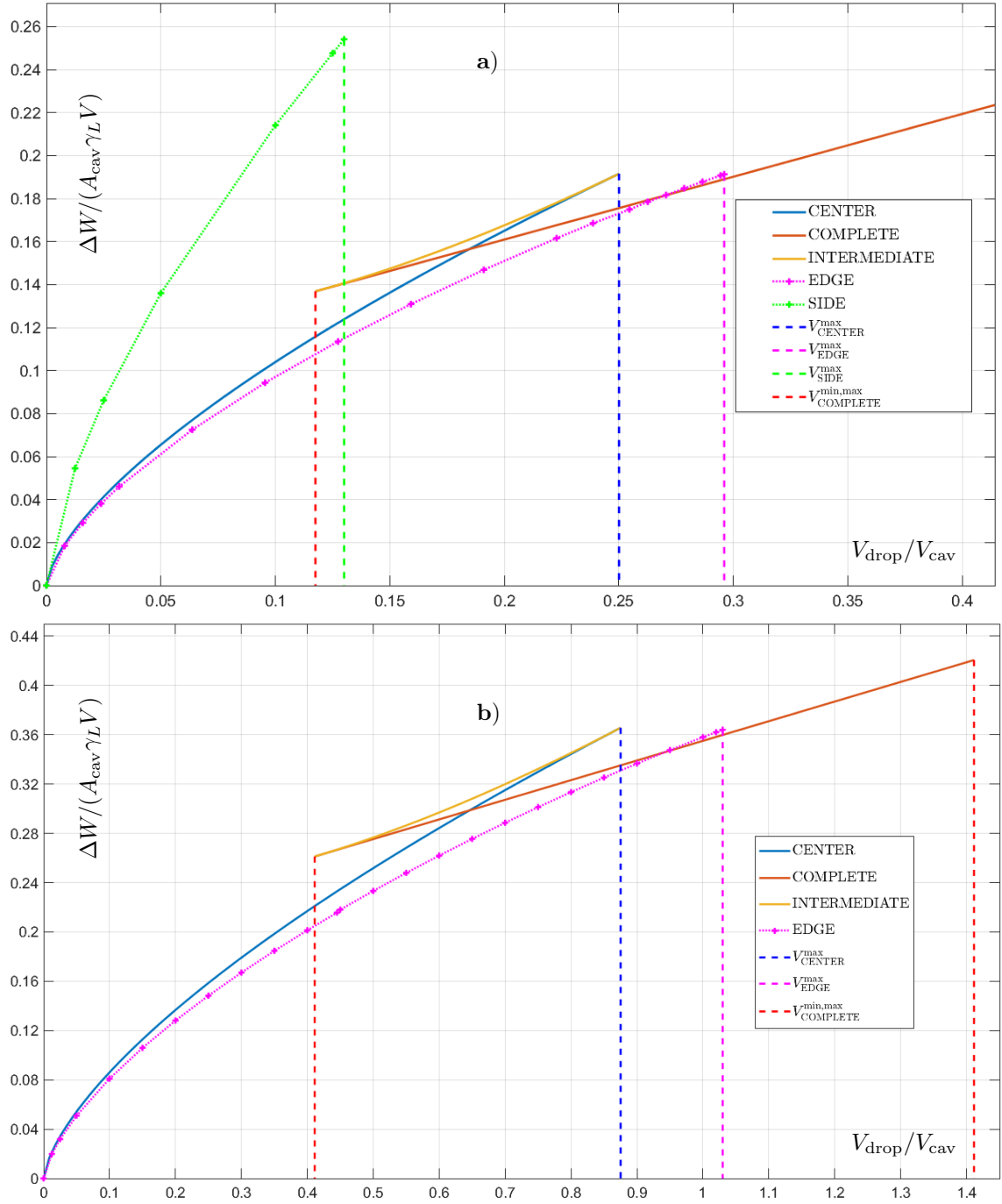

We notice that independently on the aspect ratio the EDGE configuration realizes the minimum surface energy at low volumes, while at large volumes the surface energy minimum is attained in the COMPLETE configuration. As shown in Figure \ref{fig:wsurh2-7}, when the EDGE configuration exists it always has a lower surface energy than the CENTER configuration. Then, at a volume near to the upper limit of existence (for instance, $V_{drop}/V_{cav} = 0.27$ for $R/h = 1$ and $V_{drop}/V_{cav} = 0.95$ for $R/h = 3.5$), the EDGE configuration surface energy becomes higher than the one of the COMPLETE configuration, which increases slower than the others. We also notice in Figure 3 that the INTERMEDIATE configuration always has a higher energy than the CENTER, EDGE and COMPLETE configurations. In all situations we have studied, the surface energy of the SIDE configuration is extremely high so that for the sake of simplicity, we did not represent it in Figure 3b. This is a consequence of the important contact angle of the droplet with the wall $\theta_w = 116^\circ$, so the SIDE configuration will not be considered any further.

\begin{figure}[ht!]
    \centering
    \minipage{0.43\textwidth}
    \includegraphics[trim = 1cm 1cm 2cm 0.5cm, clip=true, width=\linewidth]{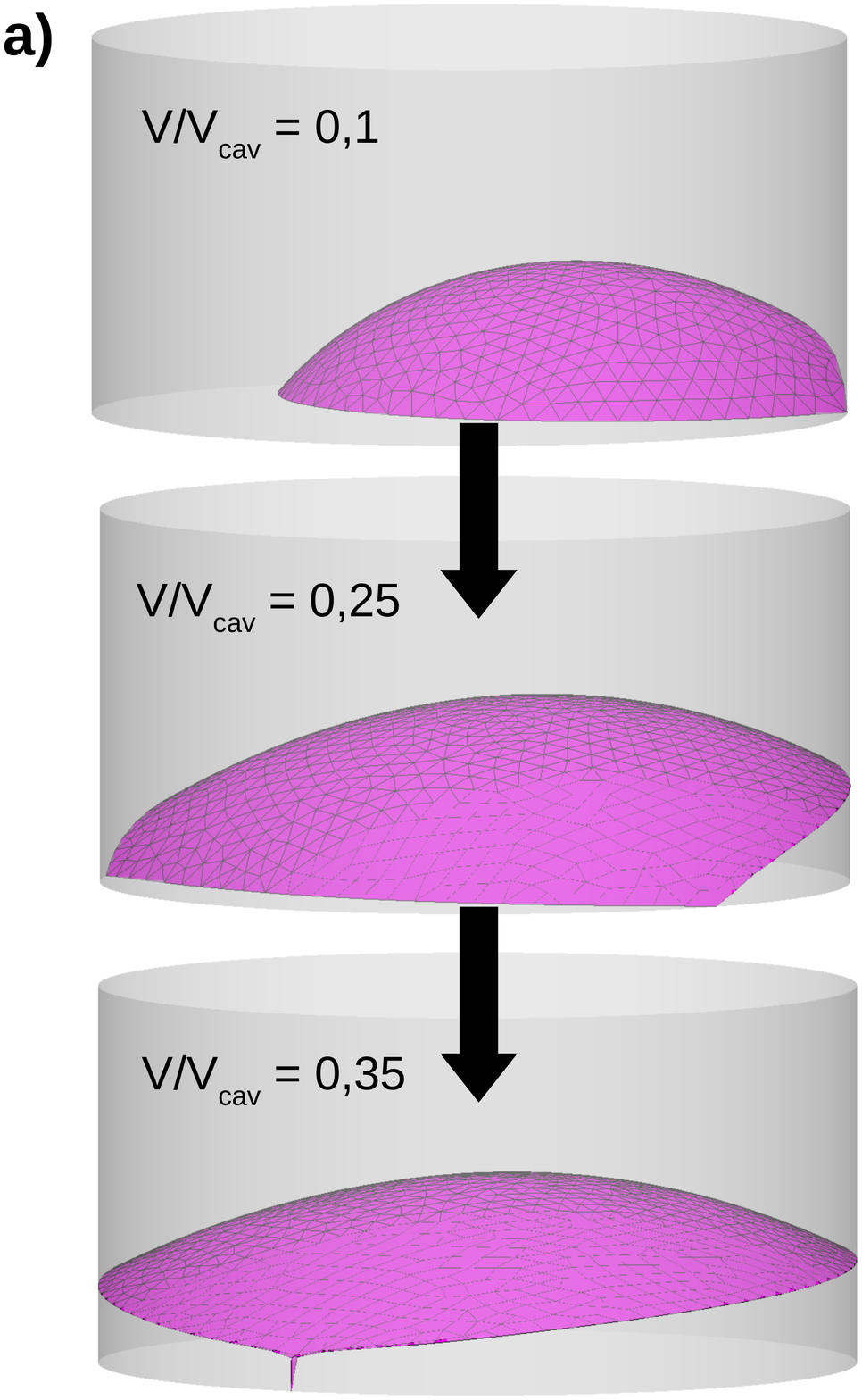}
    \endminipage
    \minipage{0.49\textwidth}
    \includegraphics[trim = 1cm 3.2cm 1cm 0.5cm, clip=true, width=\linewidth]{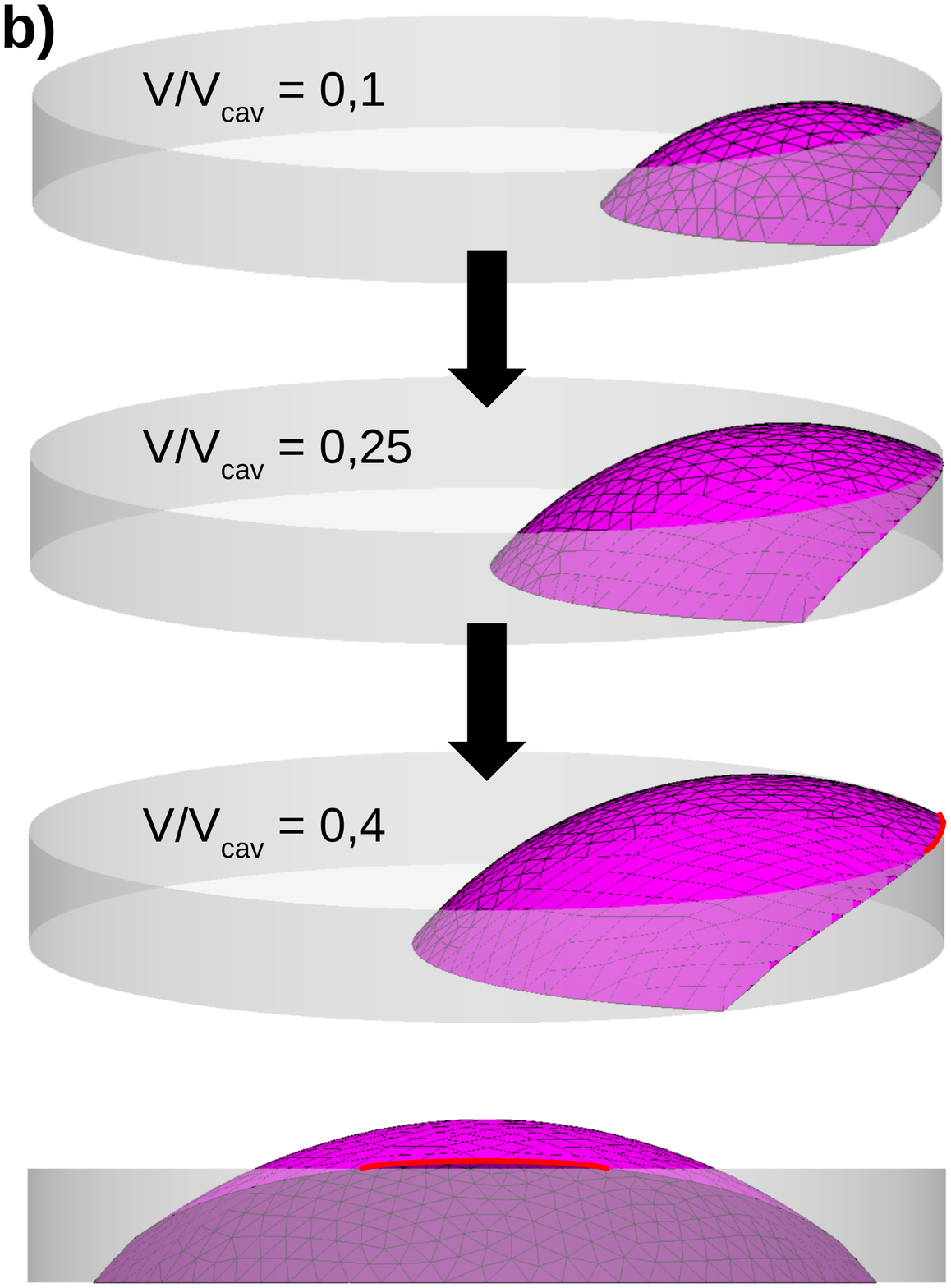}
    \endminipage
        \caption{Evolution of the EDGE configuration as a function of $V/V_{\rm cav}$: \textbf{a)} For $R/h =1$  the droplet progressively wets along the sidewall, until all the bottom of the cavity is covered. The corresponding volume is referred to as $V_{\textrm{EDGE}}^{max}$ in Figure \ref{fig:wsurh2-7}a. For $R/h=1$, $V_{\textrm{EDGE}}^{max} = 0.29$. \textbf{b)} For a high aspect ratio (here $R/h=3$) when the highest point of the droplet reach the top of the cavity at $V/V_{\rm cav}=0.35,$ the droplet will share a common boundary with the upper surface of the cavity and spread along the upper edge of the cavity until a maximum volume value (i.e., $1.03$ in Figure 3b that depends on both wetting angles $\theta_b$ and $\theta_w.$}
    \label{fig:limedge}
\end{figure}

Evolution of the EDGE configuration as a function of the droplet volume depends on the cavity aspect ratio. Figure \ref{fig:limedge} shows various droplet configurations for $R/h=1$ (left sequence) and $R/h=3.5$ (right sequence). Figure \ref{fig:wsurh2-7}a shows that in the former case the maximum volume in the EDGE configuration is attained when $V = 0.29\cdot V_{cav}$ while in the later for $V=1.03\cdot V_{cav}$ as shown in Figure \ref{fig:wsurh2-7}b.

We conclude that, as a general rule, the volume of the Ga droplet should not be too low because the EDGE configuration is always the minimum surface energy configuration at low volumes. It is interesting to notice that our results are significantly different from the ones presented in \cite{VUK19} for the same contact angles and aspect ratio $R/h=1$. In \cite{VUK19} the authors conclude\footnote{This conclusion seems to be related to the definition of the surface energy in \cite{VUK19}, which does not tends to zero when the droplet volume tends to zero.} that at very low droplet volume, the CENTER configuration has the lowest energy, then it's the EDGE configuration, and then the COMPLETE configuration. In contrast, we show here that the CENTER configuration never realizes the minimum energy morphology while the EDGE configuration (at low droplet volumes) and COMPLETE configurations (at large droplet volumes) are always preferred by the droplet. 

\subsection{Dewetting condition \label{par:dewet_cyl}}

\begin{figure}[ht!]
    \includegraphics[width=\textwidth]{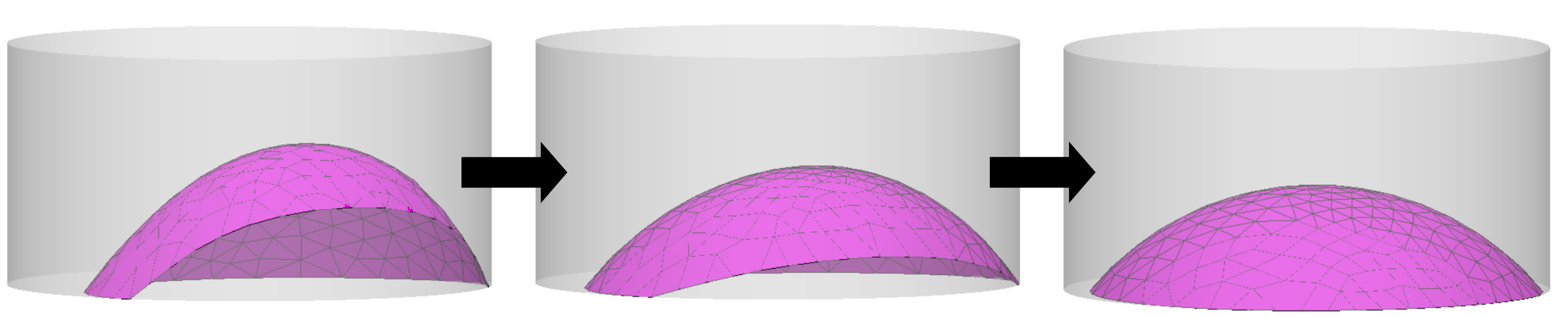}
        \caption{Configurations corresponding to different itterations with Surface Evolver showing progressive dewetting of the drop for $V/V_{cav} = 0.2$ and an aspect ratio of 2. Here, $\theta_b = 50^\circ$, $\theta_w = 150^\circ$ so (\ref{eq:dewet}) is fulfilled.}
    \label{fig:dem_cyl}
\end{figure}

In order to complete this study various values of contact angles were also tested. An interesting result was obtained when the contact angles $\theta_w$ and $\theta_b$ satisfy the relation :

\begin{equation}
    \theta_w - \theta_b > \frac{\pi}{2}.
    \label{eq:dewet}
\end{equation}
In this case, initially placed in the EDGE configuration, the droplet leaves the sidewall of the cavity after a few iterations. This is the classical "dewetting" of the sidewall represented in Figure \ref{fig:dem_cyl}. The final minimum surface energy configuration is one when the droplet wets only the bottom of the cavity which is a configuration similar to the CENTER one, except that, due to the initial condition in the minimization process, it is shifted toward the side of the cylinder base. The dewetting will be discussed further in subsection \ref{gendewet}. At first sight, this is a very interesting property for our applications as it should avoid initial tilted NW growth during the VLS process. Condition (\ref{eq:dewet}) is the same as the dewetting condition presented in \cite{microdrop} in the case of a droplet wetting two perpendicular planes.

\section{Droplet in a conical cavity}

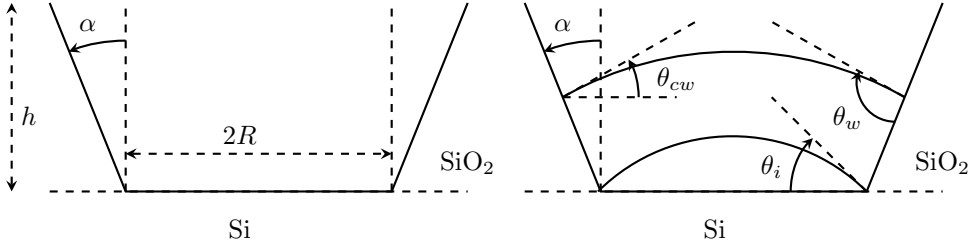
\begin{figure}
\centering
\minipage{0.45\textwidth}
\begin{tikzpicture}[scale=0.5]
%\draw[help lines] (0,0) grid (28,8);
%\draw (0,0) -- (0,2) -- (4,2) -- (4,0) -- (0,0);
\draw[thick] (1,6) -- (3,1) -- (10,1) -- (12,6); 
\draw[thick,dashed] (3,1) -- (3,6); 
\draw[thick,dashed] (10,1) -- (10,6);
\draw[thick, ->,>=stealth] (3,5) arc (90:112:4);
\draw[thick,dashed, <->,>=stealth] (3,2) -- (10,2);
\draw[thick,dashed, <->,>=stealth] (0,1) -- (0,6);
\draw (6,2) node[above] {$2R$};
\draw (0,3) node[right] {$h$};
\draw (2,5) node[above] {$\alpha$};
\draw[thick, dashed] (1,1) -- (12,1);
\draw (6,0.5) node[left,below] {Si};
\draw (12,2.3) node[left,below] {SiO$_2$};
\end{tikzpicture}
\endminipage
\minipage{0.45\textwidth}
\begin{tikzpicture}[scale=0.5]
%\draw[help lines] (0,0) grid (28,8);
%\draw (0,0) -- (0,2) -- (4,2) -- (4,0) -- (0,0);
\draw[thick] (1,6) -- (3,1) -- (10,1) -- (12,6); 
\draw[thick,dashed] (3,1) -- (3,6); 
%\draw[thick,dashed] (10,1) -- (10,6);
\draw[thick, ->,>=stealth] (3,5) arc (90:112:4);
\draw[thick, -] (10,1) arc (45:135:5);
\draw[thick, -] (11,3.5) arc (60:120:9);
\draw[thick, dashed] (10,1) -- (7.5,3.5);
\draw[thick, ->,>=stealth] (8,1) arc (180:135:2);
\draw[thick, dashed] (11,3.5) -- (7.5,5.5);
\draw[thick, ->,>=stealth] (10.75,2.85) arc (270:160:1);
\draw (9.5,3.5) node[left,below] {$\theta_w$};
\draw (7.5,2.2) node[left,below] {$\theta_i$};
\draw[thick, dashed] (2,3.5) -- (5,3.5);
\draw[thick, dashed] (2,3.5) -- (5.5,5.5);
\draw[thick, ->,>=stealth] (4,3.5) arc (0:30:2);
\draw (5,4) node {$\theta_{cw}$};
%\draw[thick, dashed] (4,3.5) ;
%\draw[thick,dashed, <->,>=stealth] (3,2) -- (10,2);
%\draw[thick,dashed, <->,>=stealth] (0,1) -- (0,6);
%\draw (6,2) node[above] {$2R$};
%\draw (0,3) node[right] {$h$};
\draw (2,5) node[above] {$\alpha$};
\draw[thick, dashed] (1,1) -- (12,1);
\draw (6,0.5) node[left,below] {Si};
\draw (12,2.3) node[left,below] {SiO$_2$};
\end{tikzpicture}
\endminipage
\caption{Left: geometric parameters in the vertical section of the truncated cone: angle $\alpha,$ basis radius $R$ and a height $h$. Right: illustration of the droplet with the angle $\theta_{cw}=\theta_w + \alpha - \frac{\pi}{2}$. }
    \label{fig:cav_cone}
\end{figure}

Although precise and complete, the previous results do not fit the experimental observations of the Ga droplets on patterned SiO$_2$/Si substrates. Indeed, for small volumes, the droplets are seen wetting only the bottom of the cavity, but not necessarily centered. This seems contradictory with the results prviously presented as the droplet is expected to wet on the edge of the cavity: i.e., a part of the bottom and a part of the side. However, is also known that the cavities in the $\textrm{SiO}_2$/Si substrate are not exactly cylindrical (see \cite{VET19}). The process of creation of the cavities involves a treatment of the surface with fluorhydric acid to remove the remaining SiO$_2$ at the bottom of the cavities. Altogether, the resulting cavities have the shape of a truncated cone with an opening angle $\alpha \simeq 28^\circ$  as shown in Figure \ref{fig:cav_cone}. Thus, in addition to the aspect ratio $R/h,$ the trucated cone cavities have an additional geometric parameter: the opening angle $\alpha.$

In this part, all the configurations are defined in the same way as for the cylindrical cavity. The main modification is that the sidewall of the cavity is not a cylinder but a truncated cone. The CENTER configuration is not modified as it is independent from the wall of the cavity. New formulas are needed for the COMPLETE configuration and the EDGE and SIDE configurations which were studied numerically. For the INTERMEDIATE configuration, the formulas are still the same but the limit angles of existence $\theta_1$ and $\theta_2$ are modified. The tilt of the sidewall introduces a new angle, further denoted $\theta_{cw}$ 
(for \textit{conical wall}), and defined as

\begin{equation}
    \theta_{cw} = \theta_w + \alpha - \frac{\pi}{2}.
\end{equation}

Thus, the new limit angles are now $\theta_1 = \min(\theta_b,\theta_{cw})$ and $\theta_2 = \max(\theta_b,\theta_{cw})$. With our values, {\em i.e.}, $\theta_b = 50^\circ, \theta_w = 116^\circ, \alpha = 28^\circ,$ (from \cite{VET19}) we obtain $\theta_{cw} = 54^\circ$ and thus $\theta_1 = \theta_b$ and $\theta_2 = \theta_{cw}$. This is different from the cylindrical cavity where $\theta_b$ was the maximal value of $\theta_i$. Now, it is the minimum value, so that we notice that in this case, the opening of the truncated cone induce an inversion of the $\theta_i$ bounds.

\subsection{Results for a Ga droplet in a conical cavity with an angle $\alpha = 28^\circ$ \label{sec:cone28}}

\begin{figure}[ht!]
\begin{tikzpicture}
    \node[anchor=south west,inner sep=0] at (0,0) {\includegraphics[scale=0.32]{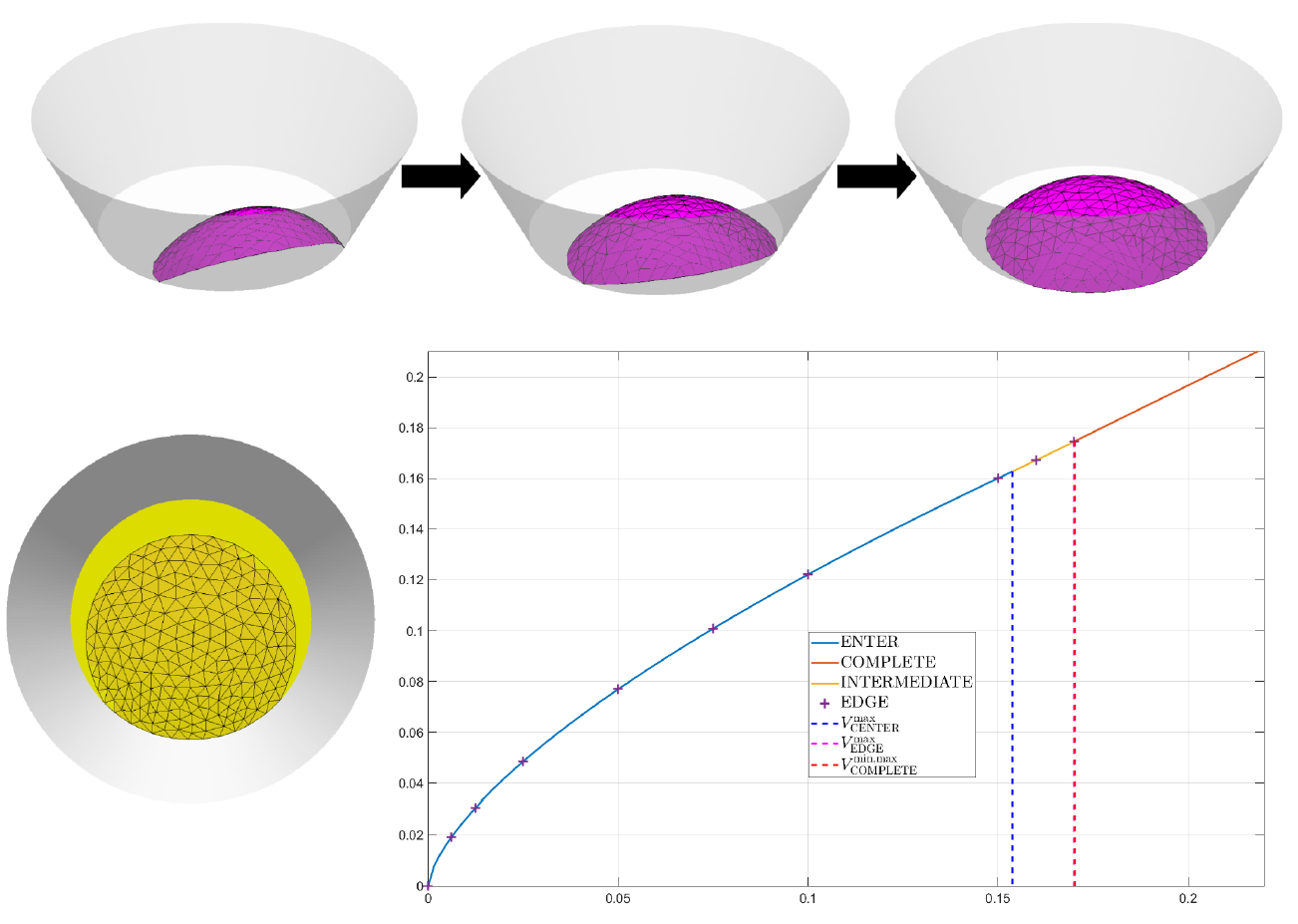}};
    \draw[black,thick,rounded corners] (0,0) rectangle (15,11);
    \draw[black,thick,rounded corners] (0,0) rectangle (4.3,7);
    \draw[black,thick,rounded corners] (0,11) rectangle (15,7);
    \draw[black,thick,rounded corners] (4.3,0) rectangle (15,7);
    \draw (0.5,10.5) node {\bf a)};
    \draw (0.5,6.5) node {\bf b)};
    \draw (4.6,6.5) node {\bf c)};
    \draw (13.5,0.8) node {$V/V_{\rm cav}$};
    \node[rotate=90] at (5.3,5){$\Delta W/(A_{\rm cav}\gamma_LV{})$};
%    \draw (1.,14) node {$V/V_{\rm cav}$};
\end{tikzpicture}
        \caption{\textbf{a)} Ga droplet dewetting the sidewall, for $R/h=1$ and $V/V_{cav} = 0.1$. \textbf{b)} Top view of the droplet once it has totally dewet the sidewall (here the droplet is transparent and in yellow we represent the bottom of the cavity; the sidewall are represented in gray). \textbf{c)} The energies of the different configurations, for aspect ratio $R/h=1$.}
    \label{fig:dem_cone}
\end{figure}
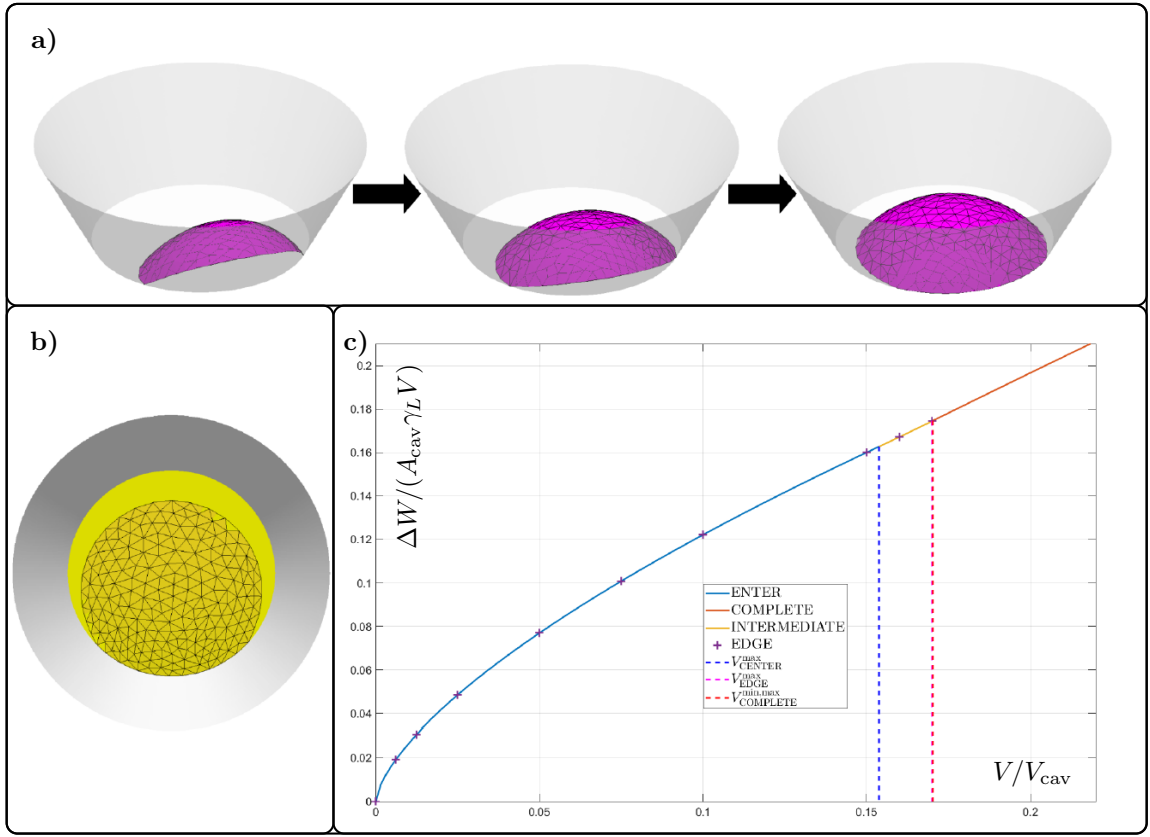

The main result concerns the evolution of a droplet initially placed in the EDGE configuration. Contrary to the results obtained for the cylindrical cavity, the contact angles of a Ga droplet wetting Si on the bottom of the cavity and SiO$_2$ on the sidewall, with $\alpha = 28^\circ$, lead to a systematic dewetting from the sidewall of the cavity. As long as the volume of the droplet is small enough not to wet all the bottom of the cavity, the droplet in the EDGE configuration always leaves the sidewall to wet only the bottom, thus being in a configuration similar to the CENTER one but not centered inside the cavity, as shown Figures \ref{fig:dem_cone}a and \ref{fig:dem_cone}b. The energies of the different configurations are shown in Figure \ref{fig:dem_cone}c (except as previously for the SIDE configuration, due to its much higher energy).

At low volumes, the energy of the EDGE configuration is equal to the one of the CENTER configuration. This is indeed coherent with the dewetting observed : the droplet only wets the bottom of the cavity with the angle $\theta_b$, being in the CENTER configuration but shifted toward the corner due to its initial position. Once $V_{\textrm{CENTER}}^{max}$ is passed, the droplet adopts the INTERMEDIATE configuration. It wets the whole bottom of the cavity and the contact angle varies from $\theta_b = 50^\circ$ to (see Figure \ref{fig:dem_cone}) $\theta_{cw} = 54^\circ$. Like smaller droplets, the EDGE droplet in this interval of volume totally dewets the sidewall of the cavity, now wetting the whole bottom as in INTERMEDIATE configuration, hence the equality of the energies noticed in the Figure \ref{fig:dem_cone}. Beyond $V_{\textrm{INTER}}^{max} = V_{\textrm{COMPLETE}}^{min}$, the EDGE configuration does not exists anymore. The volume is such that the droplet is forced to wet the sidewall because all the bottom is wet with the maximum possible contact angle.

Consequently, for a realistic conical cavity, our conclusions for the Ga droplet is that at low volumes, it wets the bottom of the cavity only, until the maximal volume which can be put in the bottom is reached. Then, the droplet adopts the only configuration possible, i.e. the COMPLETE one, wetting the whole bottom and also a part of the sidewall, symmetrically. Other similar calculations and measurements with the same angles but with different aspect ratios (not shown here) confirm this dewetting of this sidewall dewetting phenomenon, when the volume is low enough. It should be noticed that this result is actually independent from the aspect ratio of the cavity.

\subsection{Generalized dewetting condition \label{gendewet}}

In this subsection we generalize the dewetting condition obtained in subsection \ref{par:dewet_cyl} in a particular case. Taking into account the new geometric parameter of the conical cavity, i.e. the angle $\alpha$, we conclude that the dewetting from the sidewall occurs when the opening of the truncated cone, i.e. $\alpha$ is large enough. This condition reads : 

 \begin{equation}
     (\theta_w + \alpha) - \theta_b > \frac{\pi}{2}.
     \label{eq:cond_demouille_gene}
 \end{equation}

With the numerical values for a Ga droplet in a SiO$_2$/Si cavity, with $\theta_b = 50^\circ$ and $\theta_{w} = 116^\circ$, relation (\ref{eq:cond_demouille_gene}) is equivalent to :
 \begin{equation}
     \alpha > \frac{\pi}{2}+\theta_b -\theta_w  = 24^\circ.
     \label{eq:alpha_dem}
 \end{equation}

Thus for conical cavities on SiO$_2$/Si with an angle greater than $24^\circ$, we should observe a dewetting similar from the one described in subsection \ref{sec:cone28}. For the later, the dewetting condition was indeed fullfilled for $\alpha = 28^\circ > 24^\circ $. Other tests were made with the same contact angles but with a greater $\alpha$, for instance $\alpha = 40^\circ$. Accordingly to our condition, we observed systematic dewetting too, independently from the aspect ratio.

\begin{figure}[ht!]
\begin{tikzpicture}
    \node[anchor=south west,inner sep=0] at (0,0) {\includegraphics[scale=0.36]{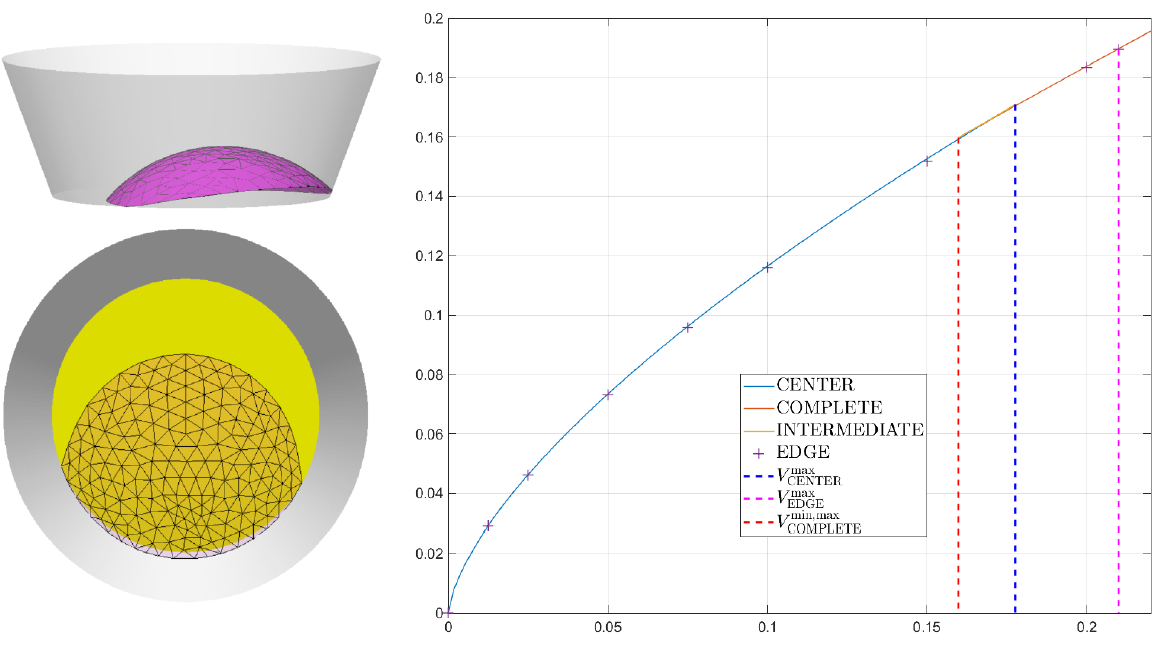}};
    \draw[black,thick,rounded corners] (0,0) rectangle (15,8.5);
    \draw[black,thick,rounded corners] (0,0) rectangle (5,5.4);
    \draw[black,thick,rounded corners] (0,5.4) rectangle (5,8.5);
    \draw[black,thick,rounded corners] (5,0) rectangle (15,8.5);
    \draw (0.5,8) node {\bf a)};
    \draw (0.5,5) node {\bf b)};
    \draw (6,7.5) node {\bf c)};
    \draw (13.7,0.6) node {$V/V_{\rm cav}$};
    \node[rotate=90] at (6.2,5){$\Delta W/(A_{\rm cav}\gamma_LV{})$};
%    \draw (1.,14) node {$V/V_{\rm cav}$};
\end{tikzpicture}
        \caption{\textbf{a)} With $\alpha = 20^\circ$, the Ga droplet in the EDGE configuration still wets a part of the sidewall, for $R/h=1$ and $V/V_{cav} = 0.1$. \textbf{b)} Top view of the same droplet (the droplet is transparent. In yellow : the bottom of the cavity; in grey : the sidewall). \textbf{c)} The energies of the different configurations, for $R/h=1$. The tiny difference between the EDGE energy and the CENTER and COMPLETE ones can be noticed on the Figure (red crosses are always lower that the blue and orange curves).}
    \label{fig:cone_20}
\end{figure}
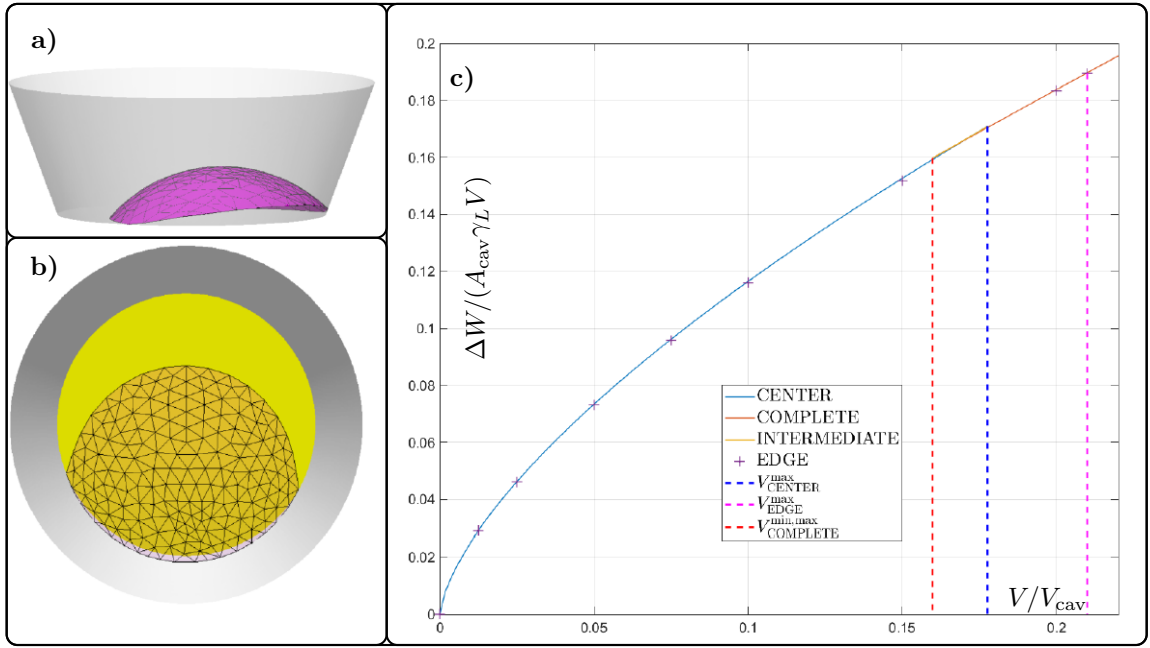

Values of $\alpha$ below $24^\circ$ were also tested : numerical results for $\alpha = 20^\circ$ are shown in Figure \ref{fig:cone_20} where the considered values of $\theta_b$ and $\theta_w$ were the same as previously. In Figure \ref{fig:cone_20}a and b the droplet in the EDGE configuration still wets a small part of the sidewall after the iterations : there is no total dewetting, as expected from relation (\ref{eq:alpha_dem}). The energy landscape shown in Figure \ref{fig:cone_20} c) shows that the energy of the EDGE configuration is very close to the one of the other configurations (because the droplet has almost left the sidewall), but still it is always slightly lower. The behavior of the droplet is then hard to predict due to the tiny difference in surface energies, but the minimum energy configuation is, as in the cylindrical case, the EDGE configuration at low volume and then at larger volumes, the COMPLETE one.

The experimental angle $\alpha = 28^\circ$ should lead to systematic dewetting from the sidewall and thus to a vertical growth, at least as long as the volume of liquid Ga is not too high. However, this angle is very close to the limit angle $24^\circ$ so that small variations around this angle (due for instance to uncertainties on the contact angles of the droplet on the bottom and on the sidewall or due to cavity defects) could explain the low proportion of tilted NWs we still observe. 

\begin{figure}[ht!]
    \centering
    \includegraphics[width=\textwidth]{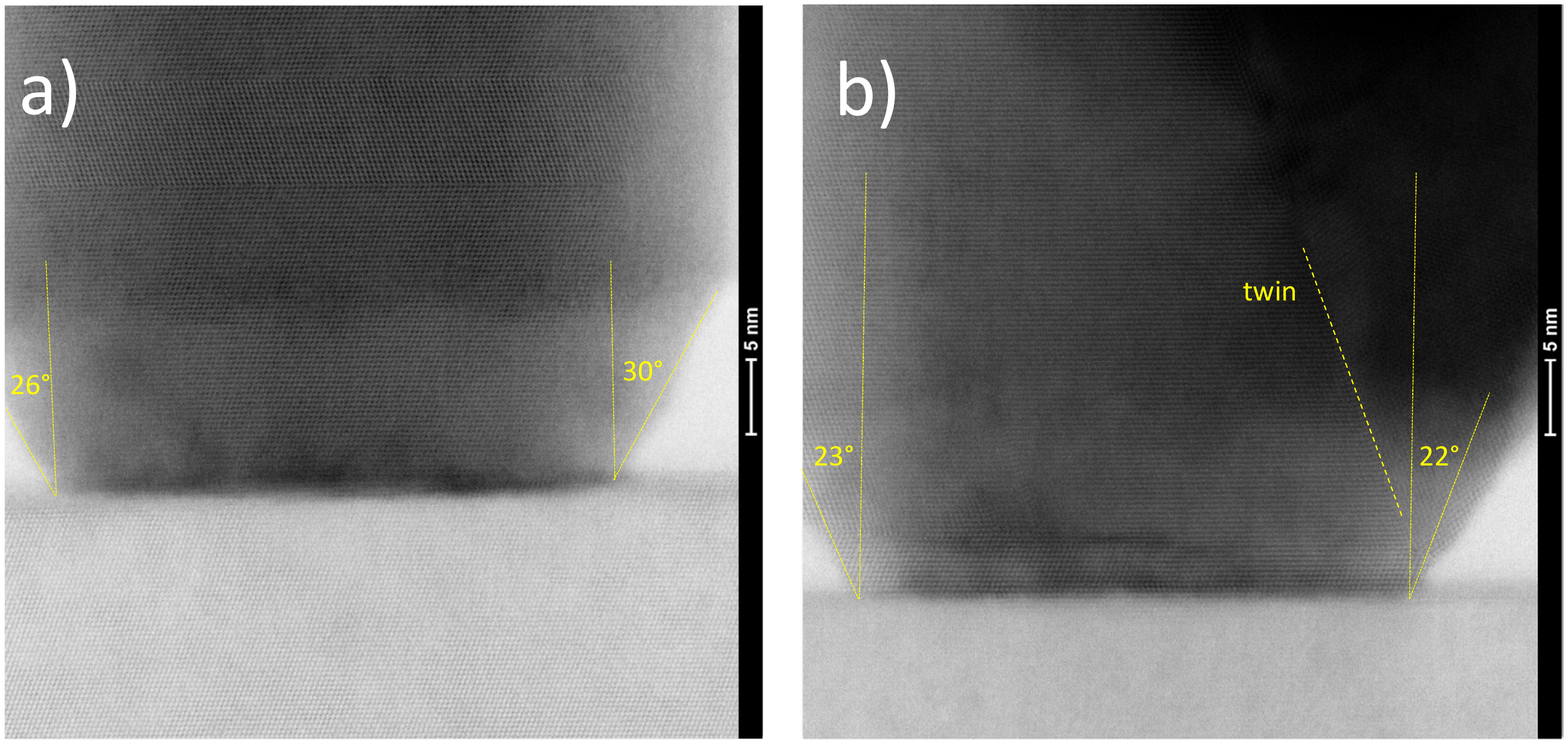}
    \vspace*{-2cm}
        \caption{\textbf{a)} Transmission electron microscopy images (with $[1\overline{1}0]$ zone axis) of GaAs NWs grown by MBE on conical cavities. In (a) the NW grew vertically due to a high value of the $\alpha$ angle (in the range $(26^\circ,30^\circ)$) while in (b) the GaAs NW didn't grew vertically due to the low value of the $\alpha$ angle here $22^\circ-23^\circ$ near the bottom of the cavity and non-constant sidewall inclination (non-uniformity). Notice that : in (a) the normal direction to the substrate is slightly inclinated (to the left) in the TEM image; in (b) the indicated twin seems to be corelated to the change in the sidewall inclination.}
    \label{fig:TEM}
\end{figure}

The modelled dewetting actually corresponds quite well to what we have observed and reported in \cite{VET19}. In those experiments, GaAs NWs were grown by Molecular Beam Epitaxy (MBE) on SiO$_2$/Si(111) patterned substrates with conical cavities in the SiO$_2$ layer (typically 20 nm-thick) fabricated by electron-beam lithography and etching (more details on the growth conditions can be found in ref \cite{VET19}). Then transmission electron microscopy images were performed on different epitaxial NWs (see Figure \ref{fig:TEM}). Interestingly, while most of NWs grew vertically (as the one visible in Figure \ref{fig:TEM}a, some of them exhibit a inclined growth direction (Figure \ref{fig:TEM}b). The angle $\alpha$ formed by the facets of the GaAs NW and the normal to the Si surface is in the range $(26^\circ,30^\circ)$ (near the bottom of the cavity) in Figure \ref{fig:TEM}a and in the range $(22^\circ,23^\circ)$ (near the bottom of the cavity) in Figure \ref{fig:TEM}b, respectively, in good agreement with our numerical estimates. 

It is also interesting to notice that experimental observations indicate that accidentally, the orientation of the lateral surface of cavities in SiO$_2$ is not constant, situation not covered by our results. Nevertheless, the above results allow to speculate that in conical cavities with non-constant sidewall inclination the Ga droplets will wet the sidewall of the cavity at different heights, so that in that case the triple line is not planar. As a consequence, a part of the triple line will attain the border of the cavity although the remaining part still wet la sidewall of the cavity leading to inclined NW growth. This scenario and our previous results point out the importance of the geometry of the cavity as well as its homogeneity in order to improve verticallity.

\section{Conclusion}

As the initial step of the nanowire growth process involves droplets in various cavities, the analysis of the minimal surface energy configurations of a droplet in a cavity becomes an important factor to control the nanowire verticality. 

We compute and compare the surface energy of morphologically different configurations for droplets wetting a cylindrical cavity in situations in which the wetting angle on the bottom of the cavity and the wetting along the sidewall are different. Depending on the aspect ratio of the cylindrical cavity, defined here as radius/heigth, some of these configurations can exist only in some specific droplet volume intervals. We find that as a general rule, the EDGE configuration realizes the minimum surface energy for small droplet volumes while the COMPLETE configuration is the minimum surface energy for large droplet volumes. An important feature of the EDGE configuration is that the triple line (contact between the droplet and the sidewall and/or bottom of the cavity) is not planar, as it seems to be also the case when the inclination of the lateral walls is not constant. In these configurations, evolution of the nanowire is such that the higher part of the triple line will reaches the top of the cavitiy while the remaining part still wet the sidewall leading to an inclinated NW growth.

From a more general perspective, depending on the values of the wetting angles on the bottom of the cavity and on the sidewall, we notices that dewetting on the sidewall of a cylindrical cavity takes place when the condition $\theta_w-\theta_b>\pi/2$ is fullfilled. Control of the dewetting the sidewall of the cavity is very important for the nanowire growth and in particular to improve the verticality. Taking the above inequality as a starting point, we extend our result to cover also the wetting of a conical cavity, in which case the geometry of the cavity has an additional geometric parameter, the opening angle $\alpha.$ We show that dewetting on the sidewall for a conical cavity occurs when $\theta_w + \alpha -\theta_b >\pi/2$ an inequality that generalizes that obtained for cylindrical cavities. 

An useful perspective of this study that we will address in a future work, is the wetting in conical cavities with orientation dependent inclination. The main difficulty in this case is the choice among all possible configurations of those pertinent to NW growth applications.     

\section*{\normalsize Acknowledgement} We acknowledge founding from the Projet ANR-15-CE05-0009 HETONAN. We also acknowledge Prof K. Brakke for his prompt answering and precious insight regarding the implementation of constraints in Surface Evolver (\cite{site}).


\begin{thebibliography}{12}

\bibitem{MAC20} P.C.McIntyre, A.Fontcuberta i Morral, Materials Today Nano 9, 100058 (2020) https://doi.org/10.1016/j.mtnano.2019.100058

\bibitem{WAG64} Wagner, R. S., Ellis, W. C. Vapor-Liquid-Solid Mechanism of Single Crystal Growth. Appl. Phys. Lett. 4, 89 (1964). https://doi.org/10.1063/1.1753975.

\bibitem{FOU19} L. Fouquat, M. Vettori, C. Botella, A. Benamrouche, J. Penuelas, G. Grenet, X-ray photoelectron spectroscopy study of Ga nanodroplet on silica terminated silicon surface for nanowire growth, Journal of Crystal Growth 514 (2019) 83–88 https://doi.org/10.1016/j.jcrysgro.2019.03.003.

\bibitem{RA17} Yong-Ho Ra, Roksana Tonny Rashid, Xianhe Liu, Jaesoong Lee, Zetian Mi, Scalable Nanowire Photonic Crystals: Molding the Light Emission of InGaN, Advanced Functional Materials Volume27, Issue 38 1702364 (2017) https://doi.org/10.1002/adfm.201702364.

\bibitem{VUK17} Jelena Vukajlovic-Plestina, Wonjong Kim, Vladimir G. Dubrovsk, Gözde Tütüncüoğlu, Maxime Lagier, Heidi Potts, Martin Friedl, Anna Fontcuberta i Morral, Nano Letters 17, 4101 (2017) https://doi.org/10.1021/acs.nanolett.7b00842

\bibitem{KIN08} Sachin Kinge, Mercedes Crego-Calama, David N. Reinhoudt, Self-Assembling Nanoparticles at Surfaces and Interfaces, ChemPhysChem Volume9, Issue1, (2008) 20-42 (2008) 10.1002/cphc.200700475. 

\bibitem{DEM21} Valeria Demontis, Valentina Zannier, Lucia Sorba, Francesco Rossella, Surface Nano-Patterning for the Bottom-Up Growth of III-V Semiconductor Nanowire Ordered Arrays, Nanomaterials 2021, 11, 2079. https://doi.org/10.3390/nano11082079.

\bibitem{VUK19} J. Vukajlovic-Plestina, W. Kim, L. Ghisalberti, G. Varnavides, G. Tütüncuoglu, H. Potts, M. Friedl, L. Güniat, W.C. Carter, V.G. Dubrovskii, A. Fontcuberta i Morral, Fundamental aspects to localize self-catalyzed III-V nanowires on silicon, Nature Communications 10, 869 (2019) https://doi.org/10.1038/s41467-019-08807-9.

\bibitem{YUA18} Yuan X. et al. Role of surface energy in nanowire growth. J. Phys. D: Appl. Phys. 51, 283002 (2018)

\bibitem{HAR18} Harmand J.-C. et al. Atomic Step Flow on a Nanofacet. Phys. Rev. Lett 19 166101 (2018) https://doi.org/10.1103/PhysRevLett.121.166101

\bibitem{MAL21} Carina B. Maliakkal, Marcus Tornberg, Daniel Jacobsson, Sebastian Lehmann, Kimberly A. Dick, Nanoscale Advances 3, 5928 (2021) DOI: 10.1039/D1NA00345C

\bibitem{VET19} M. Vettori, V. Piazza , A. Cattoni, A. Scaccabarozzi, G. Patriarche, P. Regreny, N. Chauvin, C. Botella, G. Grenet, J. Penuelas, A. Fave, M. Tchernycheva, M. Gendry, Nanotechnology 30, 084005 (2019) https://doi.org/10.1088/1361-6528/aaf3fe

\bibitem{MAT16} Matteini F. et al. Impact of the Ga Droplet Wetting, Morphology, and Pinholes on the Orientation of GaAs Nanowires. Crystal Growth \& Design. 16, 5781 (2016) https://doi.org/10.1021/acs.cgd.6b00858


\bibitem{gibbs} Gibbs, J. W., The Scientific Papers of J. Willard Gibbs: In Two Volumes; Ox Bow Press: Woodbridge, CT, 1993; Vol. 1.

\bibitem{site} Brakke K. The Surface Evolver. Disponible sur : http://facstaff.susqu.edu/brakke/evolver/evolver.html

\bibitem{SE_article} Brakke K. A. The Surface Evolver. Experimental Mathematics . 1 janvier 1992. Vol. 1, n°2, p. 141‑165.

\bibitem{manuel} Brakke K. A. Surface Evolver Manual . Version 2,70. 25 août 2013

\bibitem{microdrop} Berthier J., Brakke K. A. The Physics of Microdroplets. Hoboken (N.J.) : John Wiley \& Sons ; Salem (Mass.) : Scrivener Publishing LLC, 2012. 556 p.ISBN : 978-1-118-40133-0.


\end{thebibliography}
\end{document}